\documentclass[12pt,notitlepage,oneside]{report}

\usepackage{buetcseugthesis}

\makeindex[intoc]

\begin{document}

% Edit as needed below this line
% %%%%%%%%%%%%%%%%%%%%%%%%%%%%%%%%%%%%%%%%%%%%%%%%%%%%%%%%

% Introduction
\chapter{Introduction}\label{intro}
In this chapter we briefly discuss about motifs,what are they and why are we interested in them. Then we briefly describe the motif finding problem and various versions of it. We also give the formal description of the planted motif search for $ (l, d) $- motifs. Then we describe different approaches of motif searching algorithms and why we chose planted version of motif search. At the end of the chapter we give a short review of the mostly known approximate and exact PMS algorithm.

\section{What are Motifs}
Motif\index{Motif} means a pattern. A DNA motif is defined as a nucleic acid sequence that has some biological significance. By biological significance we mean it can be DNA binding sites for a regulatory protein that is a transcription factor. A DNA motif may be a small segment of nucleotide chain of A,T,C or G only 5 to 25 base pair long.

\subsection{Importance of Studying Motifs}
The discovery of rare event in DNA or protein sequence may lead new biological discoveries~\cite{dinh2012qpms7}. The presence of motifs is one kind of such rare events. Various biological processes such as gene expression may be controlled by motifs. Motifs are contained in regulatory regions of a genome such as promoters, enhancers, locus control regions etc~\cite{duret1997searching}. Generally, proteins known as transcription factors regulate the expression of a gene by binding to locations of motifs in regulatory regions. For example transcription factors such as TFIID, TFIIA and TFIIB usually bind to sequence 5'-TATAAA-3' in the promoter region of a gene in order to initiate its transcription. Such motifs and their locations in regulatory regions like binding sites are important and helpful to uncover the regulatory mechanism of gene expression which is very sophisticated. Motif search also has many applications in solving some crucial biological problems. For example, finding motifs in DNA sequences is very important for the determination of open reading frames, identification of gene promoter elements, location of RNA degradation signals and the identification of alternative splicing sites~\cite{pal2016efficient}.
In protein sequences, patterns have led to domain identification, location of protease cleavage sites, identification of signal peptides, protein interactions, determination of protein degradation elements, identification of protein trafficking elements etc.  So we can agree to the fact that, motif identification plays an important role in biological studies.

\subsection{Regulatory Regions and Transcription Factor Binding Sites}
DNA regions involved in transcription and transcriptional regulation are called regulatory regions (RR). Every gene contains a RR typically stretching 100-1000 bp
upstream of the transcriptional start site. A transcription factor (TF) is a protein that can bind to DNA and regulate gene expression. TFs influence gene expression by binding to a specific location in the respective genes regulatory region which is known as TFBSs or motifs. TFBSs are a part of either the promoter or enhancer region of a gene.

\section{The Motif Finding Problem}
\begin{figure}%[H]
	\centering
	\includegraphics[width=0.6\textwidth]{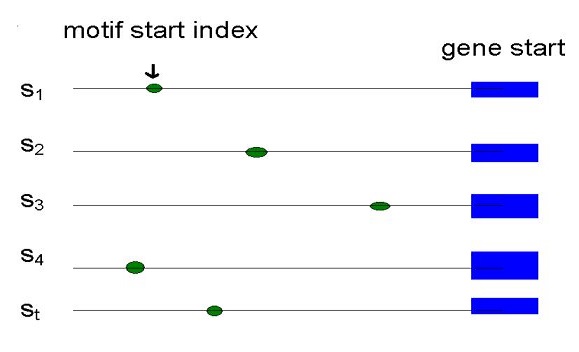}
	\caption{Starting index of motifs in a DNA sequence .}
	\label{fig:sindex}
\end{figure}

Our target is to identify transcription factor binding sites or motif from a DNA sequence. We are given a set of DNA sequences of specified length and the length of the motif. We choose a position from every sequence which will be the starting sequence for the motif candidates \Cref{fig:sindex}. Taking these sequences we make alignment matrix, profile matrix and finally consensus string. Then we evaluate the consensus string with the help of a scoring function. Now, we have to find the starting position for which the score of the consensus string will be maximum. It is the informal description of the motif searching problem. 

\subsection{Classification of Motif Searching Algorithms}
For the importance of motifs, various motif finding algorithms have been researched and applied for the last two decades. We can classify these algorithms in to three categories. All motif searching algorithms follow one of the three approaches:
\begin{itemize}
	\item \textbf{Combinatorial Approach :} Tries to explore exhaustively all the ways that a molecular process could happen thus leads to hard combinatorial problems for which efficient algorithms are required.
	\item \textbf{Probabilistic Approach :} Makes certain decisions randomly by extending the classical model of deterministic algorithms. They are often faster, simpler and more elegant than their	combinatorial counterparts. 
	\item \textbf{Phylogenetic Footprinting Approach :} Discovers regulatory elements in a set of orthologous regulatory regions from multiple species by identifying the best
	conserved motifs in those orthologous regions.
\end{itemize}

\subsection{Combinatorial Approach}
Among the various approaches combinatorial approach that has proven to be more accurate than the others. In our approach we will use combinatorial algorithms for the discovery of motifs. The are several versions of this approach. Mosty used three versions are:
\begin{itemize}
	\item Planted Motif Search (PMS)
	\item Simple Motif Search (SMS)
	\item Edited-distance-based Motif Search (EMS)~\cite{rajasekaran20091}
\end{itemize}

\subsection{Planted Motif Search (PMS)}\index{Planted Motif Search}
Among different versions of combinatorial approaches the PMS problem is more popular due to its closeness to motif reality.  Motifs typically occur with mutations at binding sites. The binding sites are referred to as instances of a motif. A motif in PMS is referred to as a $ (l, d) $-motif where $ l $ is its length and $ d $ is the
maximum number of mutations allowed for its instances. Given a set of n sequences, the PMS Algorithm tries to find all the $ (l, d) $-motifs in them. The PMS problem is essentially the same as the closest substring problem.

\subsection{Quorum Planted Motif Search (qPMS)}
A generalized version of the PMS Problem namely Quorum Planted Motif Search (qPMS) problem\index{Quorum Planted Motif Search}. The qPMS problem is to find all the motifs that have motif instances present in $ q $ out of the $ n $ input sequences. This version captures the nature of motifs more precisely than the PMS problem because in practice some motifs may not have motif instances in all of the input sequences. qPMS algorithms can be used to find DNA motifs and protein motifs as well as transcription factor binding sites. 

\subsection{Formal Definitions of qPMS}
\qquad \textbf{Definition 1 } A string $ x=x[1]\dots x[l] $ of length $ l $ is called an $ l $-mer. 

\qquad \textbf{Definition 2 } Given two strings $ x=x[1]\dots x[l] $ and $ s=s[1]\dots s[m] $ with $ l<m $ we say $ x\epsilon_{l}s $ if there exists $ 1\leq i \leq l-m+1 $ such that $ x[j]=s[l-m+1] $ for every $ 1\leq j\leq l $. We also say that $ x $ is an $ l $-mer in $ s $.

\qquad \textbf{Definition 3 } Given two strings $ x=x[1]\dots x[l] $ and $ y=y[1]\dots y[l] $ of equal length, the Hamming distance between $ x $ and $ y $ denoted by $ d_{H}(x,y) $ is the number of mismatches between them. In other words, $ d_{H}(x,y)=\Sigma_{1\leq i \leq l}I_{i} $, where $ I_{i} $ is the indicator at position $ i $. $ I_{i}=1 $ if $ x[i]\neq y[i] $ and $ I_{i}=0 $ otherwise.

\qquad \textbf{Definition 4 } Given two strings $ x $ and $ s $ with $ |x|<|s| $, the Hamming distance between $ x $ and $ s $, denoted by $ d_{H}(x,s) $ is $ \min _{y\epsilon _{|x|}s} d_{H}(x,y) $.

\qquad \textbf{Definition 5 } Given a set of $ n $ strings $ s_{1},\dots,s_{n} $ of length $ m $ each, a string $ M $ of length $ l $ is called an $ (l, d, q) $-motif of the strings if there are at least $ q $ out of the $ n $ strings such that the Hamming distance between each one of them and $ M $ is no more than $ d $.

\qquad \textbf{Definition qPMS } Given $ n $ input strings $ s_{1},\dots,s_{n} $ of length $ m $ each, three integer parameters $ l $, $ d $ and $ q $, find all the $ (l,d,q) $-motifs of the input strings. The PMS problem is a special case of the qPMS problem when $ q=n $.

\section{Literature Review}
During the last two decades various types of PMS algorithm has been proposed in the literature. There are mainly two types of algorithms for PMS problem. They are \textit{exact} and \textit{approximate} algorithms. An exact algorithm always finds all
the $ (l, d) $-motifs present in the input sequences. While, an approximate algorithm may not find all the motifs. 

\subsection{Exact PMS Algorithms}
An exact algorithm can find all the motifs in input sequences. In our research paper, we only consider exact algorithms. The exact variant of the PMS problem has been shown to be NP-hard~\cite{frances1997covering}. It means that there is no PMS algorithm that takes only polynomial time to iterate. As a result, all known exact algorithms have an exponential worst case runtime. Due to NP-hardness approximate algorithm arises for PMS. An exact PMS algorithm can be built using two approaches:

\subsubsection{Sample Driven Approach}\index{Sample Driven}
For all $ (m - l + 1)^{n} $ possible combinations of $ l $-mers coming from different strings generate the common neighborhood. Unlike pattern driven approach in \textit{Sample Driven (SD)} approaches all possible motifs generated from the $ l $-mers in input strings are of interest which could be found in polynomial time. \textit{Sample Driven} algorithms try to find comparative patterns by comparing the given length strings and looking for
local similarities between them. They are based on constructing a local multiple alignment of the given non-coding DNA sequences and then extracting the comparative patterns from the alignment by combining the segments which is common to most of the
non-coding DNA sequences.

\subsubsection{Pattern Driven Approach}\index{Pattern Driven}
For all $ |\Sigma|^{l} $ possible $ l $-mers check which are motifs. Trying all possible motif candidates take exponential space. \textit{Pattern Driven (PD)} algorithms are based on enumerating candidate patterns in a given length string and inputting substrings with high fitness. Compared with \textit{SD} algorithms, \textit{PD} algorithms can be performed intelligently so that patterns are not present in the data that are not generated.

All existing exact algorithms solve PMS problem in exponential time in some of its parameters. The most recent exact algorithms that have been proposed in the literature are Algorithm \textit{qPMS9} due to ~\cite{nicolae2015qpms9}, Algorithm \textit{PMS8} due to ~\cite{nicolae2014efficient}, Algorithm \textit{Pampa} due to ~\cite{davila2007pampa}, Algorithm \textit{qPMS7} due to ~\cite{dinh2012qpms7}, Algorithm \textit{PMSPrune} due to ~\cite{davila2007fast}, Algorithm \textit{PMS6} due to ~\cite{bandyopadhyay2013pms6mc}, Algorithm \textit{Voting} due to ~\cite{chin2005voting}, and Algorithm \textit{RISSOTO} due to ~\cite{pisanti2006risotto}.

\subsection{Approximate PMS Algorithms}
Approximate PMS algorithms usually tend to be faster than exact PMS algorithms. Typically, approximate PMS algorithms employ heuristics such as local search, Gibbs sampling, expectation optimization etc. Some examples of approximate algorithms are Algorithm \textit{MEME} due to ~\cite{bailey1994fitting}, Algorithm \textit{PROJECTION} due to ~\cite{buhler2002finding}, Algorithm \textit{Gibbs DNA} due to ~\cite{lawrence1993detecting}, Algorithm \textit{WINNOWER} due to ~\cite{pevzner2000combinatorial}, and Algorithm \textit{Random Projection} due to ~\cite{rocke1998algorithm}. Some other approximate PMS algorithms are Algorithm \textit{MULTIPROFILER} due to ~\cite{keich2002finding}, Algorithm \textit{PatternBranching} due to ~\cite{price2003finding} and Algorithm \textit{CONSENSUS} due to ~\cite{hertz1999identifying}.

\subsubsection{Algorithm MEME}\index{Algorithm!MEME}
The MEME algorithm extends the EM algorithm for identifying motifs in unaligned sequences. While a drawback of EM is that the maximum it finds is only local, MEME can either favor motifs that appear exactly once or appear zero or once in each sequence in a training set, or give no preference to a number of occurrences. In 2005 \textit{Hall et al} acquired a set of correlated genes from genomic, transcriptomic and proteomic analyses. They applied MEME to scan 1000 bp of the 3' end of stop codon, where a 47 bp motif was found in six of the analyzed sequences. Then it was used to search the entire genome and 20 additional genes were identified to have the same motif. This motif was known to be bound to Puf protein, implying that Puf protein may control the transcription of the analyzed genes.

\subsubsection{Algorithm Projection}\index{Algorithm!Projection}
This algorithm ameliorates the limitations of existing algorithms by using random projections of input. It extends previous projection-based searching techniques to solve a multiple alignment problem that is not effectively addressed by pairwise alignments. It is designed to efficiently solve the problems from the planted $ (l, d) $-motif model and can do more reliably and substantially difficult instances than previous algorithms. For $ t= 20 $ and $ n= 600 $ this algorithm achieves performance close to the best possible, being limited primarily by statistical considerations.

\subsubsection{Algorithm Winnower, SP-STAR, and cWinnower}\index{Algorithm!Winnower}
Winnower first represents motif instances as vertices then it tries to delete spurious edges and recover motifs with the remaining vertices. SP-STAR is a local sum of pairwise score improvement algorithm which considers only the subsequences present in dataset and iteratively updates scores of the motifs. cWinnower improves its running time by a stronger constraint function.

% preliminaries
\chapter{Preliminaries}\label{preliminaries}
In this chapter we have briefly described the necessary topics we need to
know before we go into Motifs. We have started with the most fundamental
part of life cells. Then from the cells, we have gone deeper into the genetic
materials the genes, DNA, RNA and Proteins. We have discussed about the flow
of information from the DNA to the Protein in detailed steps. We have illustrated the motif finding problem and classify different types of motifs. Later, we have shown building the consensus string from the alignment matrix and profile matrix. We have also discussed a consensus string evaluating function. We have concluded this chapter by giving a very simple brute-force motif finding algorithm.

\section{What is Life Made of}
In 1665 Robert Hooke discovered that every organisms
in living bodies are composed of individual compartments
known as cells\index{Cell}. With this huge discovery in
the field of Biology the study of life became the study of cells.

\subsection{Cells}
A cell is the smallest structural unit of an organism that
is capable of independent functioning. Moreover, a cell as a complex mechanical system with many moving parts which not only stores all the necessary information to make a
complete replica of itself but also contains all the machinery required to collect
and manufacture its components, carry out the copying process, and start its new offspring. So, cells are the fundamental working units of every living system.

\subsection{Life Cycle of a Cell}
\begin{figure}[!tb]
	\centering
	\includegraphics[width=0.8\textwidth]{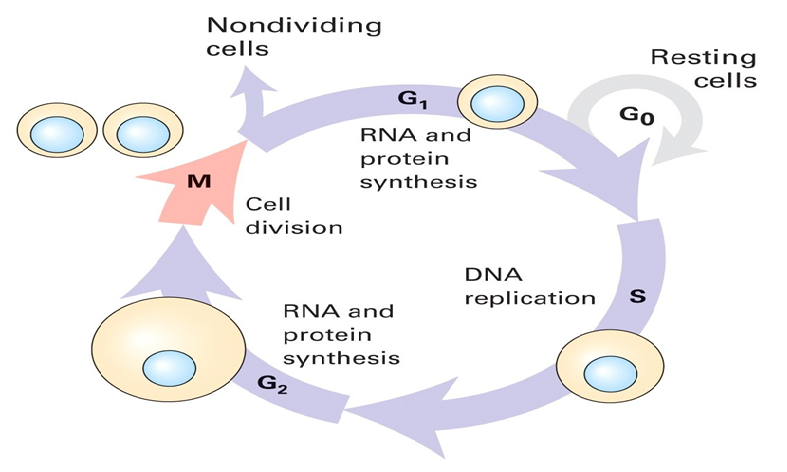}
	\caption{Life cycle of a cell.}
	\label{fig:cellcycle}
\end{figure}

A great diversity of cells exist in nature, but they all have some common
features. All cells have a life cycle: they are born, eat, replicate, and die
illustrated in \Cref{fig:cellcycle}

\subsection{Types of Cells}
We can differentiate cells into two types based on their
structures. They are called \textit{Prokaryotic} and \textit{Eukaryotic} cells.
Prokaryotic\index{Cell!Prokaryotic} cells do not contain a nucleus or any other membrane-bound organelle.
Only contain one piece of circular DNA and no mRNA post transactional modification.
Bacteria are an example of prokaryotes.
Eukaryotic\index{Cell!Eukaryotic} cells contain membrane-bound organelles, including a nucleus. 
They contain multiple chromosomes. Eukaryotes can be single-celled or multi-celled such as humans, plants and fungi.

\begin{figure}[H]
	\centering
	\includegraphics[width=0.7\textwidth]{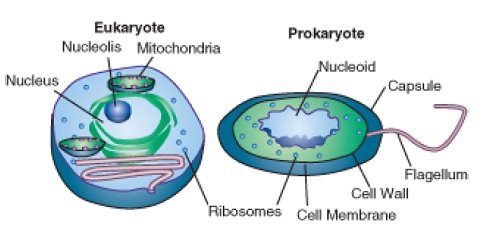}
	\caption{Prokaryotic and Eukaryotic cells.}
	\label{fig:celltype}
\end{figure}

\section{Genetic Material of Life}
All the genetic materials of a organism is called Genome\index{Genome}.
The genome is an organism’s complete set of DNA, including all of its genes. 
Each genome contains all of the information needed to build and maintain that organism.
They includes both the genes and non-coding sequences of the DNA.
A bacteria genome contains about 600,000 DNA base pairs while
human and mouse genomes have some 3 billion. Different organisms have different numbers of chromosomes, suggesting that they might carry information specific for each species. Human genome has 46 (23 pairs) distinct chromosomes. Each chromosome contains many genes.

\subsection{Genes}
Genes\index{Gene} are discrete units of hereditary information located on the chromosomes and consisting of DNA. Gregor Mendel's experiments with garden peas suggested the existence of genes that were responsible for inheritance. Genes are small sections of DNA within the genome that code for proteins. They contain the instructions for our individual characteristics- like eye and hair color. 
The human genome contains approximately 25000 protein-coding genes.
So, we can say that the main job of genes-
\begin{itemize}
	\item Storing information about the characteristics. 
	\item They express the genotype and phenotypes.
	\item The main task of genes is to produce proteins.
	\item Each gene contains the information required to build specific proteins needed in an organism.
\end{itemize}
 
\subsection{DNA}
DNA\index{DNA} or deoxyribonucleic acid, is the hereditary material in humans and almost all other organisms. Nearly every cell in a person’s body has the same DNA. DNA was discovered by Johann Friedrich Miescher by isolating a substance he called \textit{nuclein} from the nuclei of white blood cells. The information in DNA is stored as a code made up of four chemical bases: \textit{Adenine} (A), \textit{Guanine} (G), \textit{Cytosine} (C), and \textit{Thymine} (T). Human DNA consists of about 3 billion bases, and more than 99 percent of those bases are the same in all people.

DNA bases pair up with each other, A with T and C with G, to form units called base pairs. Each base is also attached to a sugar molecule and a phosphate molecule. Together, a base, sugar, and phosphate are called a nucleotide. Nucleotides are arranged in two long strands that form a spiral called a double helix. The structure of the double helix is somewhat like a ladder, with the base pairs forming the ladder's rungs and the sugar and phosphate molecules forming the vertical sidepieces of the ladder.

\begin{figure}[!tb]
	\centering
	\includegraphics[width=0.7\textwidth]{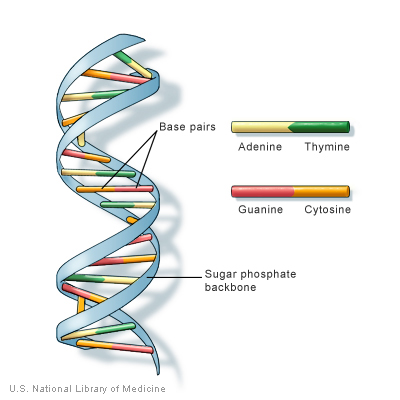}
	\caption{DNA double helix formed by base pairs attached to a sugar-phosphate backbone.}
	\label{fig:dna}
\end{figure}

Although,DNA has a double helix structure, it is not symmetric. It has a ``forward" and ``backward" direction.  The ends are labeled 5' and 3' after the Carbon atoms in the sugar component like 5' AATCGCAAT 3'. DNA always reads 5' to 3' for transcription replication. DNA is a polymer of Sugar-Phosphate-Base. Bases held together by \textit{Hydrogen} bonding to the opposite strand. Base pairs of G and C contain three hydrogen bonds and base pairs of A and T contain two hydrogen bonds. As a reason, G-C base-pairs are more stable than A-T base pairs. 

\begin{figure}[!tb]
	\centering
	\includegraphics[width=0.7\textwidth]{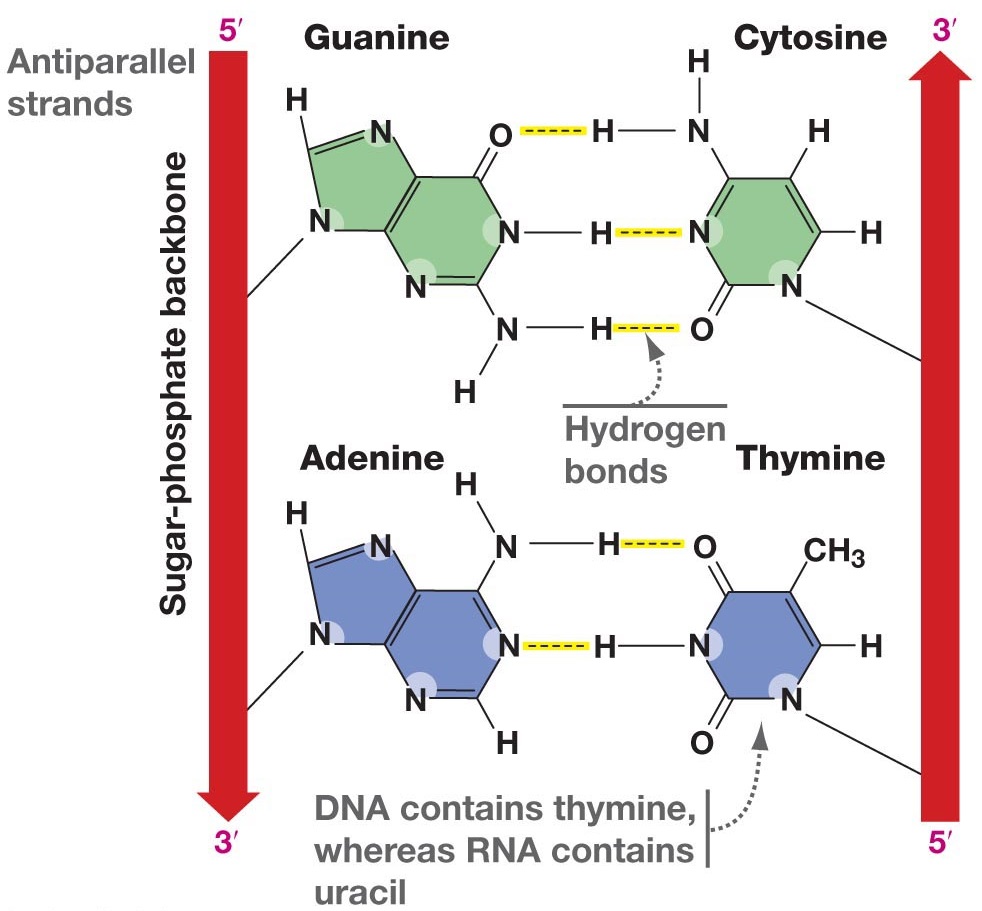}
	\caption{Hydrogen bonds between A-T and C-G.}
	\label{fig:dna2}
\end{figure}

\subsection{RNA}
RNA\index{RNA} or ribonucleic acid, is similar to DNA chemically. It usually has only a single strand. RNA contains ribose while DNA contains deoxyribose. Deoxyribose lacks one oxygen atom. RNA contains the bases \textit{Adenine} (A), \textit{Uracil} (U) instead of \textit{Thymine} in DNA, \textit{Cytosine} (C) and \textit{Guanine} (G).
Unlike DNA, RNA comes in a variety of shapes and types. While DNA contains double helix, RNA may be of more than one type. RNA is usually single-stranded, while DNA is usually double-stranded. Some forms of RNA can form secondary structures by pairing up with itself. This can have change it's properties. DNA and RNA can also pair with each other. RNA molecules are involved in protein synthesis and sometimes in the transmission of genetic information.

\begin{figure}[!tb]
	\centering
	\includegraphics[width=0.9\textwidth]{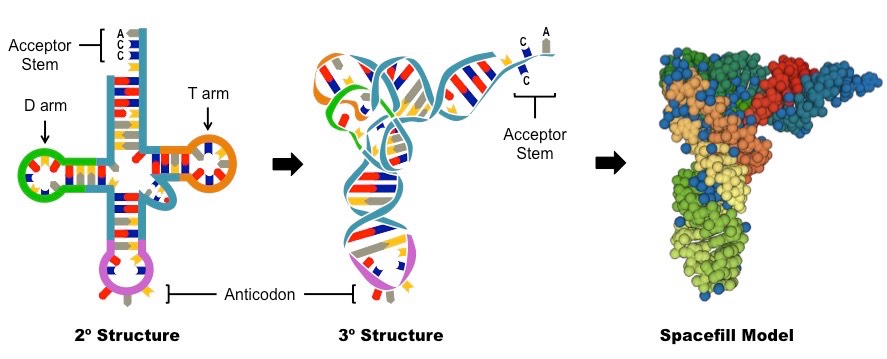}
	\caption{tRNA linear and comples structure.}
	\label{fig:rna}
\end{figure}

Several types of RNA exists. They can be classified by various functions like:
\begin{itemize}
	\item \textbf{mRNA - Messenger RNA :} Encodes amino acid sequence of a polypeptide. When a Bioinformatician says ``RNA" it usually means mRNA.
	\item \textbf{tRNA - Transfer RNA :} Brings amino acids to ribosomes during translation.
	\item \textbf{rRNA - Ribosomal RNA :} With ribosomal proteins, makes up the ribosomes, the organelles that translate the mRNA.
\end{itemize}

\subsection{Protein}
Protein\index{Protein} is a highly complex substance that is present in all living organisms. They are responsible for most of the complex functions that make life possible. They are polymer of smaller subunits called amino acids. Also called ``poly-peptides". There are twenty amino acids, each coded by three-base-sequences in DNA called ``codons" whish are degenerate. Different chemical properties cause the protein chains to fold up into specific three-dimensional structures that define their particular functions in the cell. 

\begin{figure}[!tb]
	\centering
	\includegraphics[width=0.8\textwidth]{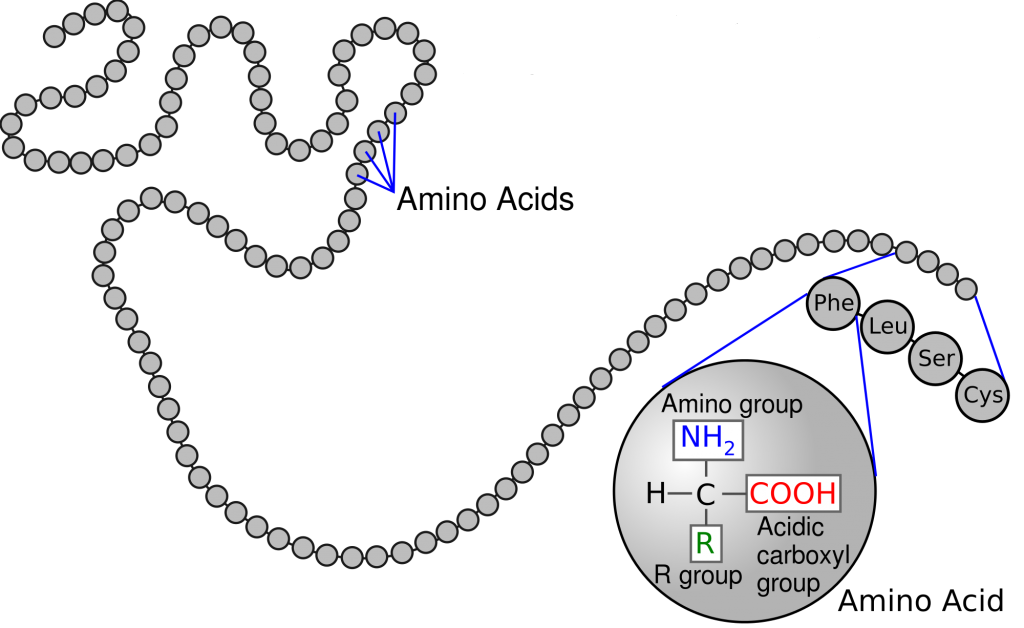}
	\caption{Primary protein structure - chain of amino acids.}
	\label{fig:protein}
\end{figure}

Proteins do all essential work for the cell like:
\begin{itemize}
	\item Build cellular structures.
	\item Digest nutrients.
	\item Execute metabolic functions.
	\item Mediate information flow within a cell and among cellular communities. 
\end{itemize}

\subsection{Mutation}
Mutation\index{Mutation} is the permanent alteration of the nucleotide sequence of the genome of an organism or DNA or other genetic elements. Mutations result from errors during DNA replication or other types of damage to DNA, which then may undergo error-prone repair or cause an error during other forms of repair or else may cause an error during replication. Mutations can serve the organism in three ways:
\begin{itemize}
	\item The Good : A mutation can cause a trait that enhances the organism’s function. For example mutation in the sickle cell gene provides resistance to malaria.
	\item The Bad : A mutation can cause a trait that is harmful, sometimes fatal to the organism. For example Huntington’s disease, a symptom of a gene mutation is a degenerative disease of the nervous system.
	\item The Silent : A mutation can simply cause no difference in the function of the organism.
\end{itemize}

\section{What Carries Information Between DNA to Proteins}
The information for making proteins is stored in DNA. There is a process (transcription and translation) by which DNA is converted to protein. By understanding this process and how it is regulated we can make predictions and models of cells.

\subsection{Central Dogma of Molecular Biology}\index{Central Dogma}
The paradigm that DNA directs its transcription to RNA, which is then translated into a protein is called central dogma of molecular biology\index{Central Dogma of Molecular Biology}. It was first proposed by Francis Crick. The central dogma of molecular biology explains the flow of genetic information, from DNA to RNA, to make a functional product, a protein.

The central dogma states that the pattern of information that occurs most frequently in our cells is:
\begin{itemize}
	\item From existing DNA to make new DNA (DNA replication).
	\item From DNA to make new RNA (transcription).
	\item From RNA to make new proteins (translation).
\end{itemize}

\begin{figure}[!tb]
	\centering
	\includegraphics[width=0.9\textwidth]{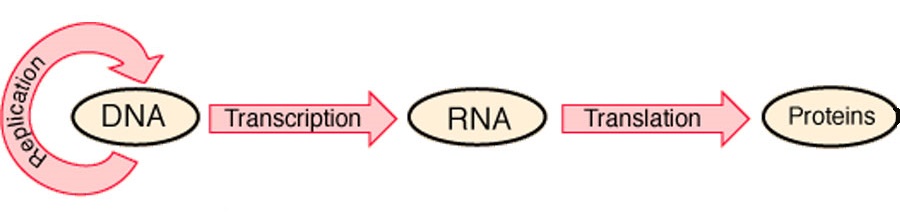}
	\caption{Central dogma of molecular biology.}
	\label{fig:dogma}
\end{figure}

\subsection{DNA$\,\to\,$RNA: Transcription}
Transcription\index{Transcription} is the process by which DNA is copied to mRNA, which carries the information needed for protein synthesis. Transcription takes place in two broad steps. First, pre-messenger RNA is formed, with the involvement of RNA polymerase enzymes. The process relies on Watson-Crick base pairing, and the resultant single strand of RNA is the reverse-complement of the original DNA sequence. The pre-messenger RNA is then edited to produce the desired mRNA molecule in a process called RNA splicing.

\subsubsection{Formation of pre-messenger RNA}
The mechanism of transcription has parallels in that of DNA replication. As with DNA replication, partial unwinding of the double helix must occur before transcription can take place, and it is the RNA polymerase enzymes that catalyze this process. Unlike DNA replication, in which both strands are copied, only one strand is transcribed. The strand that contains the gene is called the sense strand, while the complementary strand is the antisense strand. The mRNA produced in transcription is a copy of the sense strand, but it is the antisense strand that is transcribed. 

\begin{figure}[!tb]
	\centering
	\includegraphics[width=0.8\textwidth]{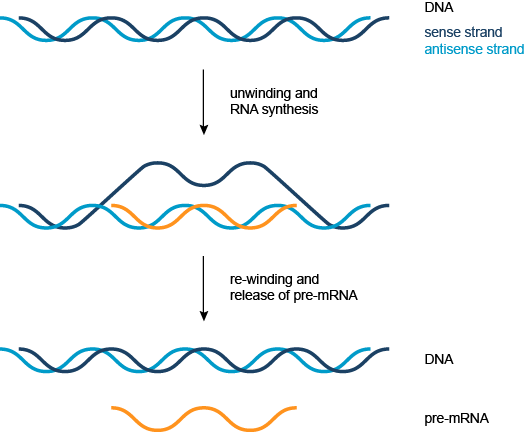}
	\caption{Formation of pre-messenger RNA.}
	\label{fig:transcription}
\end{figure}

Ribonucleotide triphosphates (NTPs) align along the antisense DNA strand, with Watson-Crick base pairing (A pairs with U). RNA polymerase joins the ribonucleotides together to form a pre-messenger RNA molecule that is complementary to a region of the antisense DNA strand. Transcription ends when the RNA polymerase enzyme reaches a triplet of bases that is read as a ``stop" signal. The DNA molecule re-winds to re-form the double helix. The \Cref{fig:transcription} illustrates the process.

\subsubsection{Splicing}\index{Splicing}
In Eukaryotic cells, RNA is processed between transcription and translation. Unprocessed RNA is composed of \textit{Introns} and \textit{Extrons}. The pre-mRNA is chopped up to remove the introns and create mRNA in a process called RNA splicing. Sometimes alternate RNA processing can lead to an alternate protein as a result. 
\begin{figure}[!tb]
	\centering
	\includegraphics[width=0.6\textwidth]{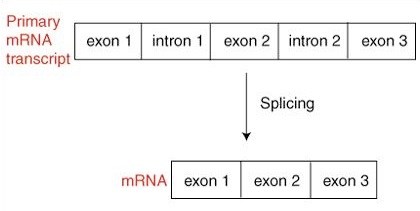}
	\caption{RNA splicing: Introns are spliced from the pre-mRNA to give mRNA.}
	\label{fig:splicing}
\end{figure}

\subsection{RNA$\,\to\,$Protein: Translation}\index{Translation}
There are twenty types of tRNAs, and twenty types of amino acids.
Each type of amino acid binds to a different tRNA, and the tRNA molecules
have a three-base segment called an anticodon that is complementary to the
``codon" in the mRNA. As in DNA base-pairing, the anticodon on the tRNA
sticks to the codon on the RNA, which makes the amino acid available to the
ribosome to add to the polypeptide chain. When one amino acid has been
added, the ribosome shifts one codon to the right, and the process repeats.
The process of turning an mRNA into a protein is called translation, since it
translates information from the RNA into the protein. 

\begin{figure}[!tb]
	\centering
	\includegraphics[width=0.6\textwidth]{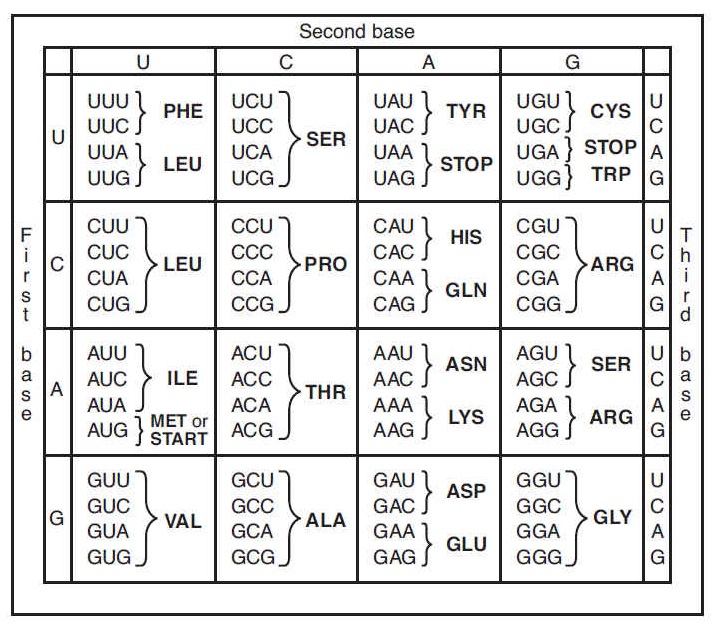}
	\caption{The genetic code, from the perspective of mRNA.}
	\label{fig:codon}
\end{figure}

%%%%%has to modify
\section{Motifs in Details}
Motif\index{Motif} usually means a pattern. For any sequence
of objects to be called a pattern, it has to have at least
more than one instance. To define more precisely, a motif
should have significantly higher occurrence in a given array
or list compared to what you would obtain in a randomized
array of same components. Here array means an arranged set
such as a genome or a protein structure. Genome is an array
of nucleotides and protein structure is an array of amino acids.

\subsection{Sequence Motifs}
By sequence motif\index{Motif!Sequence Motif} we mean a pattern
of nucleotide or amino acid that has specific biological
significance. In other words, they are pattern of a DNA array or
protein sequence. They may be also called regulatory sequence
motifs. This kind of motifs are commonly studied in bioinformatics.
They are becoming increasingly important in the analysis of gene
regulation. In our thesis we will only discuss about this type of motifs.

\subsection{Structural Motifs}
Structural motifs\index{Motif!Structural Motif} are a super-secondary
structure in a chain like biological molecules such as protein.
It is formed by three dimensional alignment of amino acids
which may not be adjacent.

\subsection{Regulatory Motifs in DNA}
Regulatory motifs\index{Regulatory Motifs} are some short nucleotide
sequence that regulates the expression of genes such as controlling
the situations under which the genes will be turned on or off.
For example:~\cite{jones2004introduction} Fruit flies have a small
set of \textit{immunity genes} that are dormant
in their genome. But when its organisms get infected somehow the genes
got switched on. It turns out that many immunity genes in
the fruit fly genome have strings that are reminiscent of TCGGGGATTTCC,
located upstream of the genes start. This string are called $ _{NF-\kappa}$B
binding sites which are important examples of regulatory motifs. Proteins
known as \textit{transcription factors} bind to these motifs, encouraging
RNA polymerase to transcribe the downstream genes.

\subsection{Regulatory Regions}
\begin{figure}[!tb]
	\centering
	\includegraphics[width=0.9\textwidth]{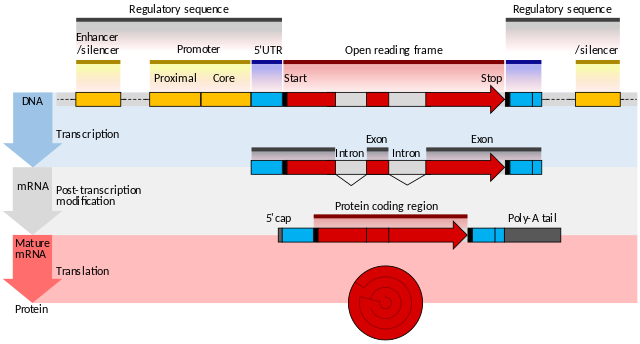}
	\caption{The structure of a eukaryotic protein-coding gene.}
	\label{fig:rr}
\end{figure}

As discussed in~\cite{Riethoven2010} regulation of gene expression
is an essential part of every organism. Certain regions,
called cis-regulatory elements, on the DNA are footprints for
the trans-acting proteins involved in transcription. DNA
regions involved in transcription and transcriptional regulation
are called regulatory regions (RR).\index{Regulatory Motifs!Regulatory Regions}
Every gene contains a RR typically
stretching 100-1000 bp upstream of the transcriptional start site.
In  \Cref{fig:rr} we see the structure of a eukaryotic protein coding gene. 
Regulatory sequence controls when and where expression
occurs for the protein coding region (red). Promoter and enhancer
regions (yellow) regulate the transcription of the gene into a
pre-mRNA which is modified to remove introns (light grey) and
add a 5' cap and ploy-A tail (dark grey). The mRNA 5' and 3'
untranslated regions (blue) regulate translation into the
final protein product.

\subsection{Transcription Factor Binding Sites}
\begin{figure}[!tb]
	\centering
	\includegraphics[width=0.9\textwidth]{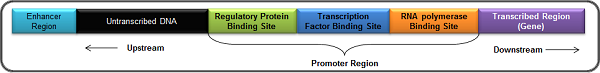}
	\caption{Transcription Factor Binding Sites.}
	\label{fig:tfbs}
\end{figure}

\index{Regulatory Motifs!Transcription Factor Binding Sites}
A transcription factor (TF) is a protein that can bind to DNA
and regulate gene expression. The region of the gene to which
TF binds is called a transcription factor binding site (TFBS).
TFs influence gene expression by binding to a specific location
in the respective gene’s regulatory region which is known as
TFBSs or motifs.~\cite{wei2007comparative} TFBS can be located
anywhere within the RR. TFBS may vary slightly across different
regulatory regions since non-essential bases could mutate.
TFBSs are a part of either the promoter or enhancer region of a gene.
A promoter sits upstream and contains three important regions-
the regulatory protein binding site, the transcription factor
binding site, and the RNA polymerase binding site.
In \Cref{fig:tfbs} we see a picture of promoter
and enhancer. An enhancer is usually far upstream of a gene.

\section{To The Motif Finding Problem}\index{Motif Finding Problem}
Our target is to identify a motif from a DNA sequence. In our study
we only deal with transcription factor binding sites. We will give
the formal definition of the motif finding problem later. Assume
that we are given some DNA sequence of nucleotides which are generated
randomly. Now, we want to find a secret pattern that occurs at least
one time in every given DNA sequence without any prior knowledge
about how it looks like. This pattern is our motif.

Some complications encountered while finding motifs:
\begin{itemize}
	\item We do not know the motif sequence.
	\item Hence we donot know what to search for in the DNA sequence.
	\item We do not know where the motifs are located relative to
	the starting index of the sequence.
	\item Motifs can differ in one or more positions in their
	sequence which is called mutations.
	\item How to discern functional motifs from random ones?
	
\end{itemize}

\begin{figure}[H]
	\centering
	\includegraphics[width=0.6\textwidth]{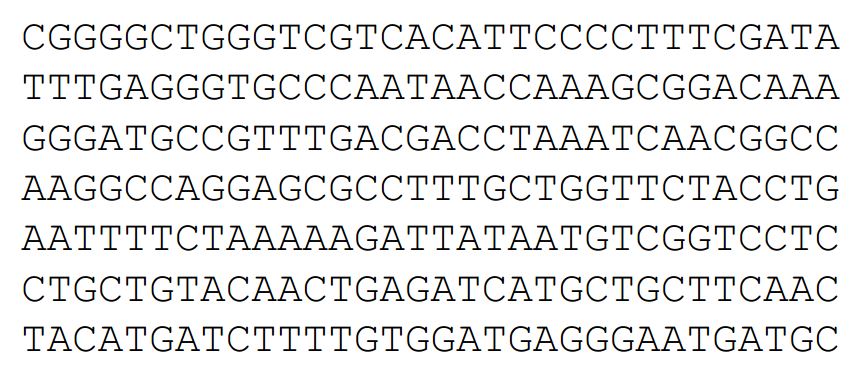}
	\caption{Seven random sequences.}
	\label{fig:seq1}
\end{figure}

\begin{figure}[!tb]
	\centering
	\includegraphics[width=0.75\textwidth]{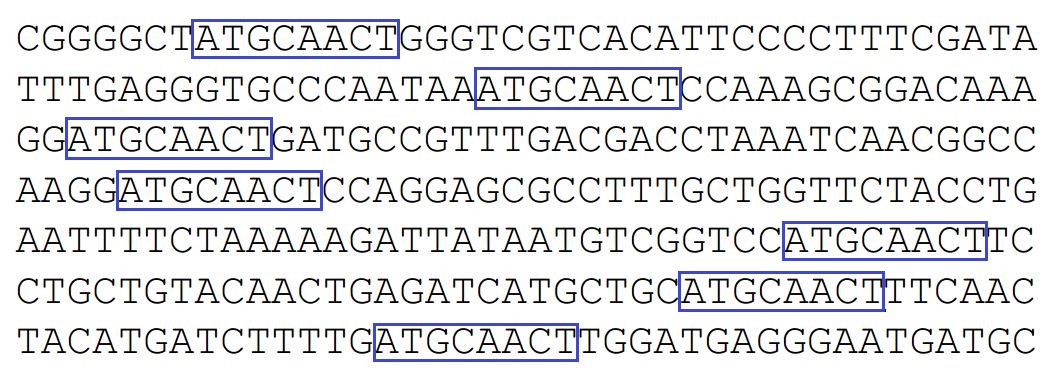}
	\caption{The same DNA sequences of \Cref{fig:seq1}
		with the implanted pattern ATGCAACT.}
	\label{fig:seq2}
\end{figure}

The idea is illustrated in the \Cref{fig:seq1}, \Cref{fig:seq2}
and \Cref{fig:seq2}.

\begin{figure}[!tb]
	\centering
	\includegraphics[width=0.75\textwidth]{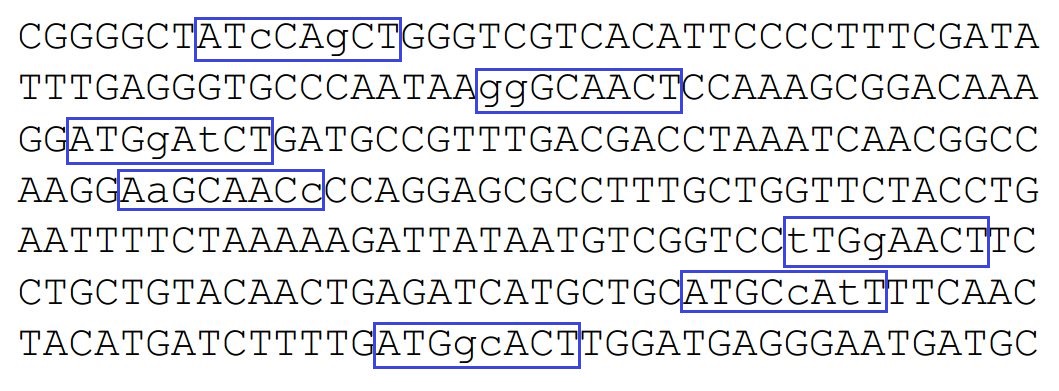}
	\caption{Same as \Cref{fig:seq2} with the implanted
		pattern ATGCAACT randomly
		mutated in two positions.}
	\label{fig:seq3}
\end{figure}

\subsection{Alignment Matrix}
\begin{figure}%[!tb]
	\centering
	\includegraphics[width=1.0\textwidth]{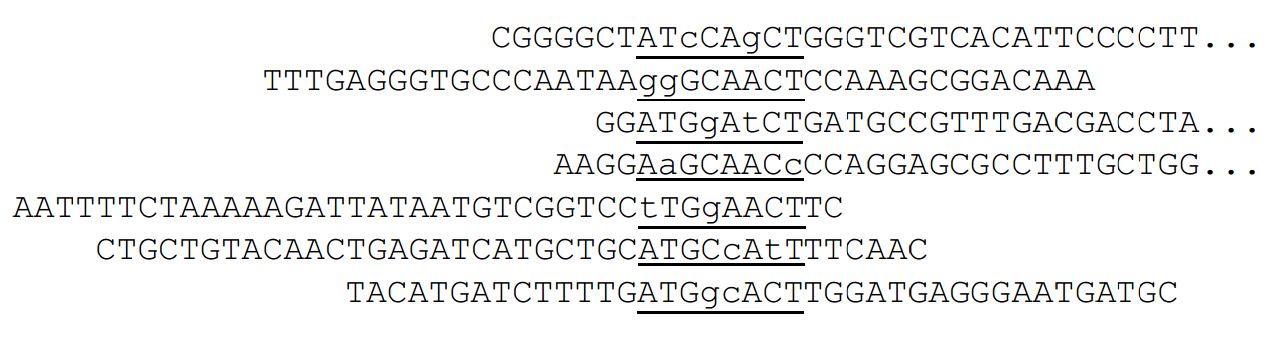}
	\caption{Superposition of the seven highlighted 8-mers from \Cref{fig:seq2}.}
	\label{fig:superpose}
\end{figure}

\index{Alignment Matrix}
To formulate the motif finding problem we first need to define
what we mean by motif. As we saw in \Cref{fig:seq3} there may be
mismatch in some positions of the motifs. Now consider a set of \textit{t} DNA sequences $ S_{1}, S_{2},\ldots, S_{t} $. Each of which has \textit{n} nucleotides. We select one position in each of these \textit{t} sequences, thus forming an array
$ s = (s_{1}, s_{2},\ldots,s_{t}) $, where $ 1 \leq s_{i} \leq n - l + 1$.
Here \textit{l} is the length of the pattern. The superposition
of the \textit{l}-mers of \Cref{fig:seq3} is shown in \Cref{fig:superpose}.
Now taking the \textit{l}-mers starting at these positions
we can compile a \textit{alignment matrix}\index{Alignment Matrix}
of $ t \times l $. The $ (i, j) $th element of the \textit{alignment matrix}
is the nucleotide at the $ s_{i}+j-1 $th position in the
\textit{i}th sequence. Illustrations are shown in \Cref{fig:consensus}.

\subsection{Profile Matrix}
\begin{figure}%[!tb]
	\centering
	\includegraphics[width=0.7\textwidth]{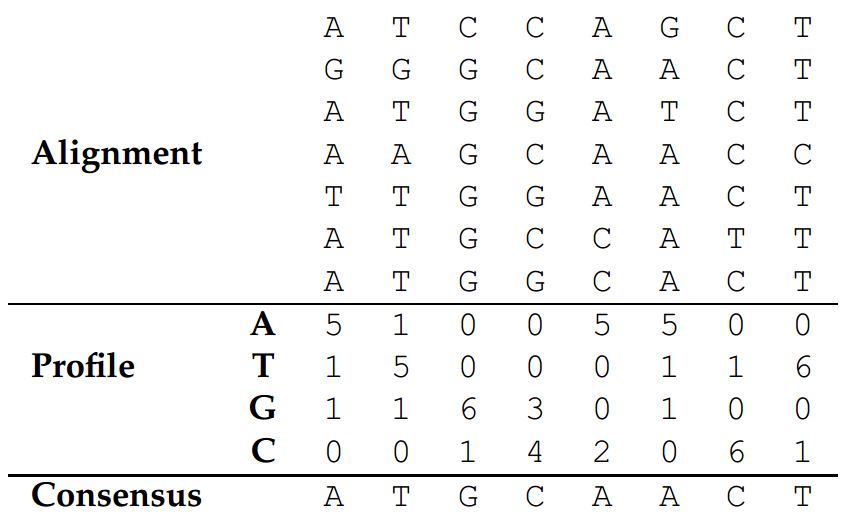}
	\caption{The alignment matrix, profile matrix and consensus
		string formed from the 8-mers starting at positions
		$ s = (8, 19, 3, 5, 31, 27, 15) $ in \Cref{fig:seq2}.}
	\label{fig:consensus}
\end{figure}

The \textit{profile matrix}\index{Profile Matrix} or profile, illustrates
the variability of nucleotide composition at each position for
a particular group of \textit{l}-mers. Let the alphabet $ \Sigma $
in our motif search. Then the size of the \textit{profile matrix} will be
$ | \Sigma | \times l$. For DNA $ | \Sigma | = 4$.  Based on the
\textit{alignment matrix}, we can compute the $ 4 \times l $
\textit{profile matrix} whose $ (i, j) $th element holds the
number of times nucleotide $ i $ appears in column
$ j $ of the \textit{alignment matrix}, where $ i $ varies from 1 to 4.
In \Cref{fig:consensus} we see that positions 3, 7, and
8 are highly conserved, while position 4 is not.

\subsection{Consensus String}\index{Consensus String}
To further summarize the profile matrix, we can form a consensus string from the most popular element in each column of the alignment matrix. It is the nucleotide with the largest entry in the profile matrix. \Cref{fig:consensus} shows the alignment matrix for s = (8,19,3,5,31,27,15), the corresponding profile matrix, and the resulting
consensus string ATGCAACT.

\subsection{Evaluating Motifs}
By varying the starting positions in $ s $, we can construct a large number of
different profile matrices from a given sample. Some profiles represent high conservation of a pattern while others represent no conservation at all. An imprecise formulation of the Motif Finding problem is to find the starting positions $ s $ corresponding to the most conserved profile. 

\subsubsection{Scoring Function}\index{Scoring Function}
If $ P(s) $ denotes the profile matrix corresponding to starting positions s, then
we will use $ M_{P(s)}(j) $ to denote the largest count in column $ j $ of $ P(s) $. For the profile P(s) in \Cref{fig:consensus}, $ M_{P(s)}(1) = 5, M_{P(s)}(2) = 5$ , and  $ M_{P(s)}(8) = 6 $. Given starting positions s, the consensus score is defined to be $ Score(s, DNA) = \sum_{j=1}^{l}M_{P(s)}(j) $. In our case, $ Score(s, DNA) =
5 + 5 + 6 + 4 + 5 + 5 + 6 + 6 = 42 $. $  Score(s, DNA) $ can be used to measure the strength of a profile corresponding to the starting positions $ s $.

\subsection{Defining The Motif Finding Problem}
the Motif Finding problem can be formulated as selecting starting positions $ s $ from the sample that maximize $ Score(s, DNA) $. 
\begin{table}[H]
	\begin{center}
		\begin{tabular}{p{300 pt}}
			\hline
			\textbf{Motif Finding Problem:}\\
			\textit{Given a set of DNA sequences, find a set of l-mers, one from each
				sequence, that maximizes the consensus score.}\\
			
			\textbf{Input:} A $ t \times n $ matrix of DNA, and l, the length of the pattern
			to find.\\
			
			\textbf{Output:} An array of t starting positions $ s = (s_{1}, s_{2},\dots, s_{t}) $ maximizing \textit{Score(s, DNA)}.\\
			\hline
		\end{tabular}
	\end{center}
\end{table}

% our work
\chapter{Our Approach}\label{ourapproach}

In this chapter we have discussed our approach to solve the generalized Quorum Planted Motif Search problem. We have proposed some increment technique on some existing algorithms for solving qPMS problem more efficiently. We have also proposed the idea to implement parallelism for solving this problem.

\section{Notations and Definitions}
We have used some common notations to describe all the algorithms described and reviewed. The notations and definitions have been summarized in table \cref{notations_and_def}. The first five notations ($\Sigma, s_i, l, d, q$) are the input data of the problem. Other notations have been used in several steps the algorithms.

%\begin{tabular}{columns}
\begin{table}
	\begin{center}
		\caption{Notations and Definitions}
		\label{notations_and_def}
		\begin{tabular}{|r|p{280 pt}|}
			\hline
			Notations & Definitions \\
			\hline
			$ \Sigma $: & The alphabet \\			
			$s_{i}$: 
			&  The input string of length $m$ over $\Sigma$ ( $1 \leq i \leq n $)\\
			$l$:
			& The motif length\\
			$d$:
			&  The maximum number of mismatches allowed in each occurrences\\
			$q$:
			& The minimum number of input strings that should have at least one occurrence.\\
			$|a|$:
			& the length of a string $a$.\\
			$a[j]$:
			&  The $j$th letter of a string $a$.\\
			$a\circ b$:
			& The string generated by concatenating two strings $a$ and $b$\\
			$d_{H}(a,b)$:
			& The Hamming distance between two strings $a$ and $b$ of the same length. It is given by the number of positions $j$ such that $a[j] \ne b[j]$\\
			$s\_{ij}^{l}$:
			& The $l$-length substring of an input string $s\_{i}$ that starts from the $j$th position.\\
			
			$S\_{i}$: 			
			& The set of all the $l$-length substrings of an input string
			$s_{i}$. $S_{i}=\{s_{i1}^{l},...,s_{i,m-l+1}^{l}\}$.
			\\
			$S_{i}^{d}$:			
			& The set of all the $(l+1)$ length substrings of an input string $s_{i}$ defined by $S_{i}^{d} = \{s_{l+1}^{i0},...,s_{i, m-l+1}^{l+1}\}$. Here, $s_{i0}^{l+1}[1] = s_{i,m - l+1}^{l+1}[l+1]$ = $\emptyset$, and $d_{H}$($\emptyset$, $\alpha$) = $\infty$ is assumed for any $\alpha$ $\epsilon$ $\Sigma$.
			\\
			$\mathcal{B}(a, R)$:
			& The set of strings in the sphere of radius $R$ centered at $a$. $\mathcal{B}(a,R)$ = \{b$\vert$ $\lvert$ b$\rvert$ = $|a|, d_{H}(a,b)\leq R\}$
			
			\\
			$n_{\mathcal{B}}(l,d)$:
			& $\lvert \mathcal{B}(a,d)\rvert$ for an $l$-length string $a$.
			\\
			$x|_{P}$:
			& The substring of $x$ composed by sequencing the letters of $x$ at the positions in a vector $P$. For example, $x|_{P}$ = GCA for $x$ = ACCGAT and $P = (4,2,5)$.\\
			$P(j)$:
			& The $j$th element of a vector $P$.\\
			$P_{+}(j)$:
 			& The $j$-dimensional vector composed of the first $j$ elements of a vector $P$.\\
 			$P_{-}(j)$:
 			& The $(|P|-j)$-dimensional vector composed of the last $|P|-j$ elements of a vector $P$.\\
 			$R_1(a,b,c)$:
 			& The set of indices $j$ satisfying $a[j] = b[j] = c[j]$.\\
 			$R_2(a,b,c)$:
 			& The set of indices $j$ satisfying $a[j] = b[j] \neq c[j]$.\\
 			$R_3(a,b,c)$:
 			& The set of indices $j$ satisfying $c[j] = a[j] \neq b[j]$.\\
 			$R_4(a,b,c)$:
 			& The set of indices $j$ satisfying $b[j] = c[j] \neq a[j]$.\\
 			$R_5(a,b,c)$:
 			& The set of indices $j$ satisfying $a[j] \neq b[j], b[j] \neq c[j]$ and $c[j] \neq a[j]$.
 			\\ \hline
		\end{tabular}
	\end{center}
\end{table}

\section{Previous Works}
There are several exact algorithms to solve the planted motif search problem. For the sake of completeness we will describe the previous works which have motivated us to develop our approach. In this section we will review qPMSPruneI \cite{davila2007fast}, qPMS7 \cite{dinh2012qpms7}, Traver String Ref \cite{tanaka2014improved} and PMS8 \cite{nicolae2014efficient}.

\subsection{qPMSPrune}
Algorithm qPMSPrune\index{qPMSPrune} for the qPMS problem was proposed by \cite{davila2007fast}. Algorithm qPMSPrune introduces a tree structure to find the possible motif candidates. Then it uses branch-and-bound technique to reduce the search space and find the expected ($l,d$) motifs for the given set of inputs. Algorithm qPMSPrune uses the d-neighborhood concept to build the tree structure. The key points of the algorithm is described as follows.

\subsubsection{Key Steps of qPMSPruneI}
Most of the motif search algorithms combine a sample diven approach with a pattern driven approach. Here in the sample driven part the $d$-neighborhood ($\mathcal{B}(x_0,d)$) of a given input string $x_0$ is generated as a tree. Every $l$-mer in the neighborhood is a candidate motif. \\

The qPMSPrune algorithm is based on the following observation.\\

\textbf{Observation 3.1} \textit{Let M be any ($l,d,q$)-motif of the input strings $s_1,\dots,s_n$. Then there exists an $i$ (with $1\leq i\leq n-q+1$) and a $l$-mer $x\epsilon s_i^l$ such that M is in $\mathcal{B(x,d)}$ and M is a ($l,d,q-1$)-motif of the input strings excluding $s_i$.}

\begin{figure}[!tb]
	\centering
	\includegraphics[width=0.9\textwidth]{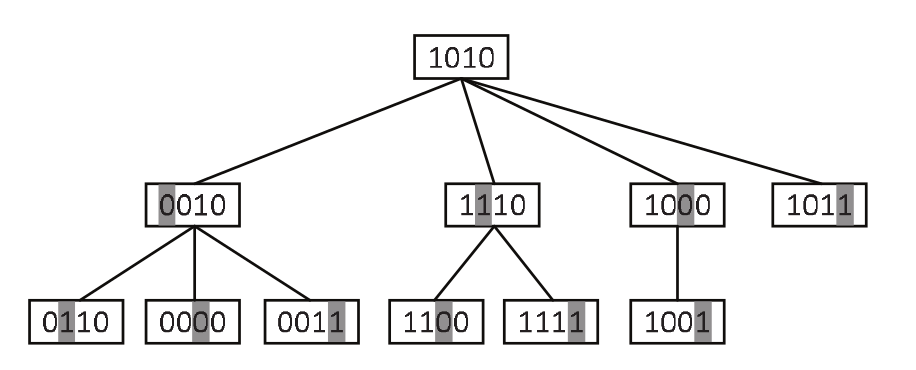}
	\caption{$\mathcal{T}_{2}(1010)$ with alphabet $\Sigma =\{0,1\}$}
	\label{fig:pmsprunetree}
\end{figure}

Based on the above observation, for any $l$-mer $x$, it represents $\mathcal{B}(x,d)$ as a tree $\mathcal{T}_{d}(x)$. It generates the $d$-neighborhood of every $l$-mer $x$ in $s_1$ by generating the tree $\mathcal{T}_{d}(x)$. The height of $\mathcal{T}_{d}(x)$ is $d$. The root of this tree will have $x$. If a node $t$ is parent of a node $t'$ then the hamming distance between the strings is $d_{H}(t,t')=1$. As a result, if a node is at level $h$, then its hamming distance from the root is $h$.\\
For example, the tree $\mathcal{T}_{2}(1010)$ with alphabet $\Sigma=\{0,1\}$ is illustrated in Figure \cref{fig:pmsprunetree}.
\\

The trees are generated and explored in depth first manner. In pattern driven part, each node is checked whether it is a valid motif or not. While traversing a node $t$ in the tree $\mathcal{T}_{d}(x)$ in a depth-first manner, the hamming distance between the input strings are calculated incrementally. If $q'$ is the number of input strings $s_j$ such that $d_{H}(t,s_j)\leq d$. If $q'\geq q-1$, output $t$ as a motif. Moreover the algorithm prunes the branch if $q" < q-1$ where $q"$ is the number of input strings $s_j$ such that $d_H(t,s_j) \leq 2d-d_{H}(t,x)$.\\

The time and space complexities of algorithm qPMSPrune are given by $O((n-q+1)nm^{2}n_{\mathcal{B}}(l,d))$ and $O(nm^2)$  respectively. The pseudocode of qPMSPrune is shown in algorithm \cref{qpmsprune}.

\begin{algorithm}[H]
	\caption{qPMSPrune}
	\label{qpmsprune}
	\begin{algorithmic}[1]
		\input{algorithms/qPMSPrune.alg}
	\end{algorithmic}
\end{algorithm}

\begin{algorithm}[H]
	\caption{qPMSPrune\_Tree($k,x_{k},p_{k},\mathcal{T}$)}
	\label{qpmsprune_tree}
	\begin{algorithmic}[1]
		\input{algorithms/qPMSPrune_Tree.alg}
	\end{algorithmic}
\end{algorithm}

\begin{algorithm}[H]
	\caption{FeasibleOccurrences2($k,x_{k},\mathcal{Q}$)}
	\label{feasibleoccurrences2}
	\begin{algorithmic}[1]
		\input{algorithms/FeasibleOccurrences2.alg}
	\end{algorithmic}
\end{algorithm}

\begin{algorithm}[H]
	\caption{IsMotif($x,q',\mathcal{T}$)}
	\label{ismotif}
	\begin{algorithmic}[1]
		\input{algorithms/IsMotif.alg}
	\end{algorithmic}
\end{algorithm}

%\begin{enumerate}
%	\item Each node in $\mathcal{T}_{d}(x)$ is a pair ($t,p$) where $t=t[1]\dots t[l]$ is an $l$-mer and $p$ is an integer between 0 and $l$ such that $t[p+1]\dots t[l] = x[p+1]\dots x[l]$. A node ($t,p$) is referred to as a $l$-mer $t$ if $p$ is clear.
%	\item Let $t=t[1]\dots t[l]$ and $t'=t'[1]\dots t'[l]$. A node ($t,p$) is the parent of a node ($t',p'$) 
%\end{enumerate}

\subsection{qPMS7}
Algorithm qPMS7\index{qPMS7} was proposed by \cite{dinh2012qpms7} which was based on qPMSPrune \cite{davila2007fast}. Some speedup techniques were introduced to improve the runtime of Algorithm qPMSPrune. 
qPMS7 also searches for motifs by traversing trees. The primary difference from qPMSPrune is that it utilizes $r\epsilon \mathcal{S}_2$ as well as $x_0 \epsilon \mathcal{S}_1$ to traverse a tree. The algorithm is based on the following observations.

\textbf{Observation 3.2} Let $M$ be any ($l, d, q$)-motif of the input strings $s_1,\dots,s_n$. Then there exist $1\leq i\neq j\leq n$ and $l$-mer $x\epsilon s_{i}^{l}$ and $l$-mer $y\epsilon s_{j}^{l}$ such that $M$ is in $\mathcal{B}(x,d) \cup \mathcal{B}(y,d)$ and $M$ is a ($l, d, q-2$)-motif of the input strings excluding $s_i$ and $s_j$.\\
Using an argument similar to the one in \cite{davila2007fast}, we infer that it is enough to consider every pair of input strings $s_i$ and $s_j$ with $1\leq i,j\leq(n-q+2)$. As a result, the above observation gets strengthened as follows.\\
\textbf{Observation 3.3} Let {$M$ be any ($l,d,q$)-motif of the input strings
$s_1,\dots,s_n$. Then there exist $1\leq i\neq j\leq n-q+2$ and $l$-mer $x\epsilon s_{i}^{l}$ and $l$-mer $y\epsilon s_{j}^{l}$ such that $M$ is in $\mathcal{B}(x,d) \cup \mathcal{B}(y,d)$ and $M$ is a ($l, d, q-2$)-motif of the input strings excluding $s_i$ and $s_j$.\\

Algorithm qPMS7 uses a routine based on the above observations which finds all of the possible motifs. Algorithm qPMSPrune explores $\mathcal{B}(x,d)$ by traversing the tree $\mathcal{T}_{d}(x)$. In Algorithm qPMS7, $\mathcal{B}(x,d)\cup \mathcal{B}(y,d)$ is explored by traversing an acyclic graph, denoted as $\mathcal{G}_{d}(x,y)$ with similar constructing rule as $\mathcal{T}_{d}(x)$.

The time and space complexity of Algorithm qPMS7 are $O((n-q+1)^{2}nm^{2}
n_{\mathcal{B}}(x,d))$ and $O(nm^2)$ respectively. So in worst case scenario the time runtime of qPMS7 is worse than that of Algorithm qPMSPrune by a factor of $n-q+1$. However, Algorithm qPMS7 is much faster than Algorithm qPMSPrune in practice.

\begin{figure}[!tb]
	\centering
	\includegraphics[width=0.9\textwidth]{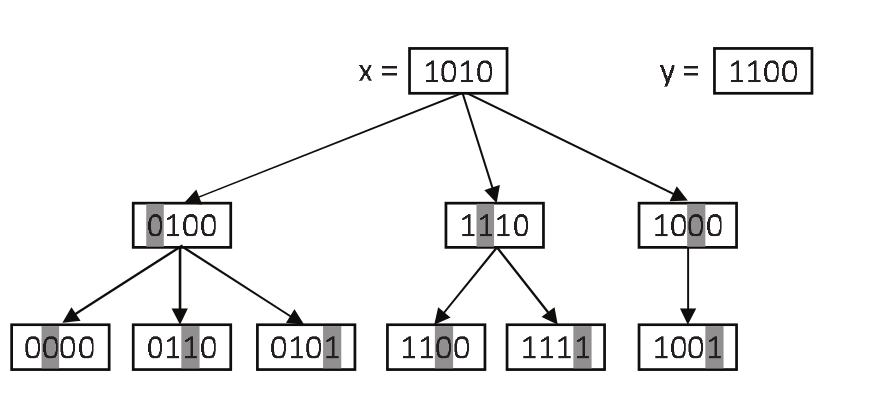}
	\caption{$\mathcal{G}_{2}(1010,1100)$ with alphabet $\Sigma =\{0,1\}$}
	\label{fig:qpms7}
\end{figure}

\begin{algorithm}[H]
	\caption{qPMS7}
	\label{qpms7}
	\begin{algorithmic}[1]
		\input{algorithms/qPMS7.alg}
	\end{algorithmic}
\end{algorithm}

\begin{algorithm}[H]
	\caption{qPMS7\_Tree($k,x_{0},r,x_k,p_k,\mathcal{T}$)}
	\label{qpms7_tree}
	\begin{algorithmic}[1]
		\input{algorithms/qPMS7_Tree.alg}
	\end{algorithmic}
\end{algorithm}

\begin{algorithm}[H]
	\caption{FeasibleOccurrences3($x_0,r,x_k,p_k,\mathcal{Q}$)}
	\label{feasibleoccurrences3}
	\begin{algorithmic}[1]
		\input{algorithms/FeasibleOccurrences3.alg}
	\end{algorithmic}
\end{algorithm}

\subsection{TraverStringRef}
TraverStringRef\index{TraverStringRef} is an improved version of qPMS7. Four improvements were proposed for which will be explained below: \newline
	
\subsubsection{Feasibility Check without Precomputed Table}
Like qPMS7 feasibility check is not performed by checking an occurrnce in the precomputed table. Here a theorem is followed:
Three strings $a,b,c$ of the same length satisfy
\begin{center}
	$\mathcal{B}(a,d_{a})\cap \mathcal{B}(b,d_{b})\cap \mathcal{B}(c,d_{c})\neq \emptyset$
\end{center}
if and only if 
\begin{center}
	$d_{a} \geq 0, d_{b} \geq 0, d_{c} \geq 0, \newline d_{a}+d_{b} \geq d_{H}(a,b), \newline d_{b}+d_{c} \geq d_{H}(b,c), \newline d_{c}+d_{a} \geq d_{H}(c,a), \newline d_{a}+d_{b}+d_{c} \geq |R_{2}(a,b,c)|+ |R_{3}(a,b,c)|+ |R_{4}(a,b,c)|+ |R_{5}(a,b,c)|$
	
\end{center}

\subsubsection{Elimination of unnecessary Combinations}
Unnecessary combinations are eliminated in qPMS7 and the procedure is quite same as qPMSPruneI. The unnecessary checks are suppressed when $q<n$ and $\mathcal{T} \leftarrow \{S_{h}|i_{2}+1 \leq h \leq n\}$.

\subsubsection{String Reordering}
This improvement is ensured by pruning the subtree rooted at current node as early as possible by checking the feasibility occurences in non decreasing order. Here the difference is when the input string from where the input reference occurences are taken is determined. Since the number of string in $\mathcal{B}(x_{0},d) \cap \mathcal{B}(r,d)$ is a non decreasing function of $d_{H}(x_{0},r)$ the reference $r$ is taken from the input that maximizes minimum hammimg distance between $x_{0}$ and $r$ to reduce size of the search tree.

\subsubsection{Position Reordering}
To ensure efficient pruning of subtrees the search tree's structure is investigated. Only the nodes satisfying the condition of not changing the hamming distance change the structure of the search tree. To take the advantage of this procedure the position of the strings reordered so that the positions where the letters are different come earlier. For this an $l$-dimensional vector is computed for each pair at the root node of the search tree.\\

The pseudo-code of the algorithm TraverStringRef is given in Algorithm [\cref{traver_string_ref}].

\begin{algorithm}[H]
	\caption{TraverStringRef}
	\label{traver_string_ref}
	\begin{algorithmic}[1]
		\input{algorithms/TraverStringRef.alg}
	\end{algorithmic}
\end{algorithm}

\begin{algorithm}[H]
	\caption{TraverStringRef\_Tree($k,x_0,r,x_k,p_k,\mathcal{T},J$)}
	\label{traver_string_ref_tree}
	\begin{algorithmic}[1]
		\STATE $\mathcal{T'} \leftarrow \emptyset$
\STATE missed $\gets$ $0$

\FORALL{$Q$ $\epsilon$ $T$}
	\STATE $Q' \leftarrow$ FeasibleOccurrencesReordered3($k,x_0,r,x_{k},p_k,Q,J$)
	\IF{$Q' \neq \emptyset$}
		\STATE $\mathcal{T} \leftarrow \mathcal{T'} \cup \{Q'\}$
	\ELSE
		\STATE missed $\gets$ missed$+$ 1
		\IF{ missed $>$ $\lvert \mathcal{T} \rvert -q + 2$}
			\RETURN
		\ENDIF
	\ENDIF
\ENDFOR

\IF{IsMotifFast($x_{k},q-2,\mathcal{T'}$) $=$ \TRUE}
	\STATE $\mathcal{M}$ $\leftarrow$ $\mathcal{M}$ $\cup$ $\{x_{k}\}$
\ENDIF

\IF{$k$ $=$ $d$}
	\RETURN
\ENDIF

\IF{$|\mathcal{T'}|$ $<$ $q-2$}
	\RETURN
\ENDIF

\STATE Sort the elements of $\mathcal{T'}$ in the nonincreasing order of $\min_{y\epsilon \mathcal{Q}} d_{H}(x_k,y)$(20)	\COMMENT{The reference needs to be corrected}

\FOR {$p_{k+1}$ $=$ $p_{k}+1$ \TO $l$}
	\FORALL{$\alpha$ $\epsilon$ $\Sigma$ $\backslash$ $\{x_{k}[J(p_{k+1})]\}$}
		\STATE $x_{k+1}$ $\gets$ $x_k$
		\STATE $x_{k+1}[J(p_{k+1})]$ $\gets$ $\alpha$
		\STATE $d_0$ $\gets$ $d+d_H(x_{0}|_{J_{-}(p_{k+1})},r|_{J_{-}(p_{k+1})})$
		
		\IF{$d_H(x_{k+1},r)$ $\leq$ min($d_0,2d-k-1)$}
			\STATE TraverStringRef\_Tree($k+1,x_0,r,x_{k+1},p_{k+1},\mathcal{T'},J$)
		\ENDIF
	\ENDFOR
\ENDFOR

	\end{algorithmic}
\end{algorithm}

\subsection{PMS8}
The key concepts of increasing efficiency in PMS8\index{PMS8} from qPMS7 and qPMSPrune are described below:

\subsubsection{Sort rows by size}
Sorted rows speed up the filtering step. As for sorted rows we need less tuples for lower stack size, it helps to reduce expensive filtering. It is because fewer $l$-mers remain to be filtered when the stack size increases.

\subsubsection{Compress $l$-mers}
To calculate the hamming distance between two $l$-mers at first we have to perform exclusive or of their compressed representation. One compressed $l$-mer requires l $\times \lceil\log|\varSigma|\rceil$ bits of storage. So the table of compressed $l$-mers only requires O(n(m-l+1)) words of memory as we only need the first 16 bit of this representation.

\subsubsection{Preprocess distances for pairs of $l$-mers}
For every pair of $l$-mers we test if the distance is no more than 2d in advance.

\subsubsection{Cache locality}
As a certain row of an updated matrix is the subset of that row so we can store the updated row in the previous location of the row. so we have to keep track the number of elements belongs to the new row. This process can be repeated in every step of recursion and can be perform using by only a single stack where the subset elements are in contiguous position of memories and thus cache locality sustains.

\subsubsection{Find motifs for a subset of strings}
we will find the motif for some input strings and then test them against the remaining strings.

\subsubsection{Memory and Runtime}
since we store all the matrices in the place of a single matrix they only require $O(n(n-l+1))$ words of memory. $O(n^2)$ words of row size will be added for at most n matrices which share the same space. So total bits of $l$-mer pair takes $O((n(m-l+1)^2)/w)$ words where w is the number of bits in a machine word. So total memory used for this algorithm is $O(n(n-l+1)+(n(m-l+1)^2)/w)$.

\subsubsection{Parallel implementation}
If we think dividing the problem into $m-l+1$ subproblems then the number of subproblems will be embarrassingly greater for parallelization. So a fixed number of subproblems are assigned to each processor. Scheduler then spawns a separate worker thread to avoid use of a processor just for scheduling. The scheduler loops until all the subproblems are solved and after the completion of all threads the motifs are given to scheduler and it outputs the result.

%\subsubsection{Pruning conditions}
%For two $l$-mers a and b the pruning condition is given below: \newline a and b have a common neighbor M such that $H_{d}(a,M) \leq d_{a}$ and $H_{d}(b,M) \leq d_{b}$ if and only if $H_{d}(a,b) \leq d_{a}+d_{b}.$

\section{Our Proposal qPMS-Sigma}
In this section we propose an algorithm qPMS-Sigma with some techniques to optimize the space complexity as well as the runtime to solve the Quorum Planted Motif Search problem. Our work is mainly based on the Algorithm TraverStringRef. 

\subsection{String Compression}\index{Compression}
We can compress the input strings $s_1,\dots s_n$ as a part of preprocessing of the algorithm. For our work we have compressed all the $l$-mers of the input strings into some group of unsigned 32-bit integers. Thus we make the motif matrix using the compressed $l$-mers. To be specific for an alphabet set $\Sigma$ we compressed a $l$-mer into $\lceil \log_{2}(\lvert \Sigma \rvert)\rceil$ groups. Each of this group contains $\lceil l/32\rceil$ 32-integers. \\

For example, for a DNA string of 32 characters, we need $\lceil \log_{2}(\lvert \Sigma \rvert)\rceil=2$ groups of unsigned 32-bit integers,  where each group contains $\lceil n/32\rceil = 1$ integers.\\

We can calculate the hamming distance between two compressed $l$-mers in two steps. First for each group of integers find out the bitwise XOR values of the integers of both the $l$-mers. After that calculate the bitwise OR values among the XOR-ed values of each group. The number of set bits in the last set of integers is the hamming distance between the two strings.\\

An example of comparing two compressed $l$-mers is given in \cref{strexample}.

\begin{figure}[H]
	\centering
	\includegraphics[width=0.9\textwidth]{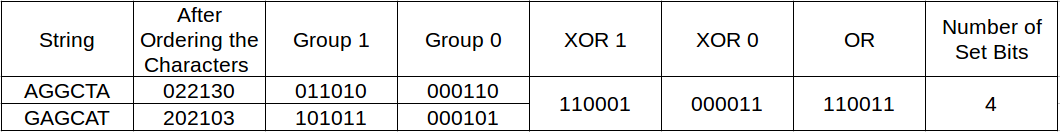}
	\caption{$l$-mer Compression Example}
	\label{strexample}
\end{figure}

%\begin{table}[H]
%	\begin{center}
%		\caption{Compression Example}
%		\label{strexample}
%		\begin{tabular}{|c|c|c|c|c|c|c|c|}
%			\hline
%			String & After Ordering the Characters & $Group_{1}$ & $Group_{0}$ & $XOR_{1}$ & $XOR_{0}$ & OR & SetBits\\
%			\hline
%			AGGCTA & 022130 & 011010 & 000110 & 110001 & 000011 & 110011 & 4 \\
%			\hline
%			GAGCAT & 202103 & 101011 & 000101 & & & & \\ \hline
%		\end{tabular}
%	\end{center}
%\end{table}

A code snippet \cref{string_compress} is written in C++ describing the structure of compressed $l$-mer and the method of calculating hamming distance.\\

%After compressing the input strings we can calculate the hamming distance between two l-mers in $O(l/(w/\log_{2}\Sigma))$, where $W$ is the size of the integer, which is 32 in our experiment.

\subsection{Motif Finding}
After compressing the input sequences we need to find the motifs in them. For these we've modified the TraverStringRef algorithm by using the compressed $l$-mers. Then we have followed the steps of Algorithm TraverStringRef \index{TraverStringRef} to find out the motifs.

\subsection{Parallel Implementation of qPMS-Sigma}\index{Parallel}
The Algorithm qPMS-Sigma can easily implement in parallel by traversing several search trees in parallel. Similar to the idea of parallelism mentioned in Algorithm TraverStringRef \cite{tanaka2014improved} we can divide the problem into $(m-l+1)$ subproblem where each subproblem explores different search trees.

\section{Space and Runtime Complexity}
Here the space complexity\index{space complexity} of our algorithm is $O(nml(\log \lceil \Sigma \rceil)/w)$. Here $w$ is the size of the integers that contain the compressed  input sequences as well as the $l$-mers. For DNA sequences $\log \lvert \Sigma \rvert = 2$. So the space complexity becomes $O(nml/w)$.\\

The worst case time complexity is similar to TraverStringRef $O((n-q+1)^2 nm^2 (m+\log n)n_\mathcal{B} (l,d))$. However, finding the hamming distance between two $l$-mers take about constant time for DNA sequence. Normally bitwise operations are a little faster than the other arithmetic operations. So, practically our algorithm runs a little faster than the former.

\section{Experimental Results}
We have compared the runtime of Algorithm qPMS-Sigma in the challenging states with the other well known algorithms. The proposed algorithm is implemented in C++ (using GNU C++ compiler). We have run the experiment on Ubuntu 14.04 (64 bit) operating system. The machine configuration is Intel\textregistered Core\texttrademark i3-4005U CPU @ 1.70GHz × 4, 4GB RAM.

The test dataset was generated in computational experiment of Algorithm TraverStringRef \cite{tanaka2014improved}. The testing dataset is randomly generated according to the FM (fixed number of mutation) model \cite{pevzner2000combinatorial}. We have used it in our experiment and showed the result in \cref{dataq20} and \cref{dataq10}. Our algorithm shows a little better result than Algorithm TraverStringRef in some challenging
states. Although it lags behind the qPMS9 in all cases. The comparisons are shown in \cref{dataq20}, \cref{dataq10}, \cref{fig:chart1}, \cref{fig:chart2}.

\begin{table}
	\begin{center}
		\caption{Parameter Setting for Testing Data Set}
		\label{dataset_parameter}
		\begin{tabular}{|ll|}
			\hline 
			Parameter
			& Setting\\
			\hline
			$|\Sigma|$
			& 4 (DNA)\\
			$n$
			& 20 \\
			$m$
			& 600 \\
			$q$
			& 10,20\\
			$l$
			& 13, 15, 17,\dots \\
			\hline
		\end{tabular}
	\end{center}
\end{table}

\begin{table}
	\begin{center}
		\caption{Computational Results for DNA Sequences for $q=20$}
		\label{dataq20}
		\begin{tabular}{|c|c|c|c|c|c|c|}
			\hline 
			\textbf{Algorithm} & \textbf{(13,4)} & \textbf{(15,5)} & \textbf{(17,6)} & \textbf{(19,7)} & \textbf{(21,8)} & \textbf{(23,9)} \\
			\hline
			\textbf{qPMS-Sigma} & 12s & 50s & 165s & 755s & 2909s & 12153s \\
			\hline
			\textbf{TraverStringRef} & 11s & 47s & 179s & 745s & 3098s & 13027s \\
			\hline
			\textbf{qPMS7} & 14s & 133s & 1046s & 8465s & 71749s & \\ \hline
			\textbf{PMS8} & 7s & 48s & 312s & 1596s & 5904s & 19728s\\ \hline
			\textbf{qPMS9} & 6s & 34s & 162s & 804s & 2724s & 8136s \\ \hline
		\end{tabular}
	\end{center}
\end{table}
\begin{figure}[!tb]
	\centering
	\includegraphics[width=0.9\textwidth]{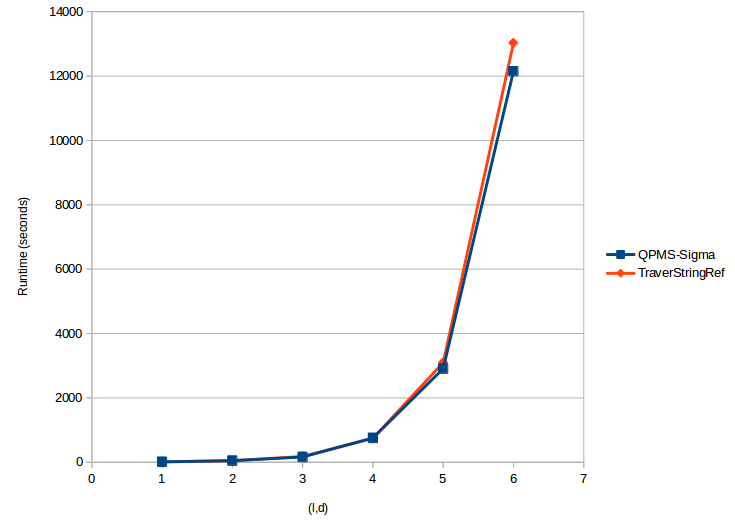}
	\caption{Run Time Comparison between qPMS-Sigma vs TraverStringRef in the challenging states for q=20}
	\label{fig:chart1}
\end{figure}

\begin{table}
	\begin{center}
		\caption{Computational Results for DNA Sequences for $q=10$}
		\label{dataq10}
		\begin{tabular}{|c|c|c|c|c|c|c|}
			\hline 
			\textbf{Algorithm} & \textbf{(13,4)} & \textbf{(15,5)} & \textbf{(17,6)} & \textbf{(19,7)} & \textbf{(21,8)} & \textbf{(23,9)} \\
			\hline
			\textbf{TraverStringRef} & 7 & 36 & 160 & 760.6 & 3643.3 & 17232.4 \\
			\hline
			\textbf{TraverStringRef} & 5.5 & 32 & 166.1 & 779.9 & 3700.6 & 17922.2 \\
			\hline
			\textbf{qPMS7} & 31 & 152 & 850.7 & & 4865.0 & 29358 \\ \hline
			\textbf{qPMSPrune} & 12.8 & 126.7 & 1116.5 & 10540.8 & & \\ \hline
			\textbf{PMS9} & & & & & & \\ \hline			
		\end{tabular}
	\end{center}
\end{table}

\begin{figure}[!tb]
	\centering
	\includegraphics[width=0.9\textwidth]{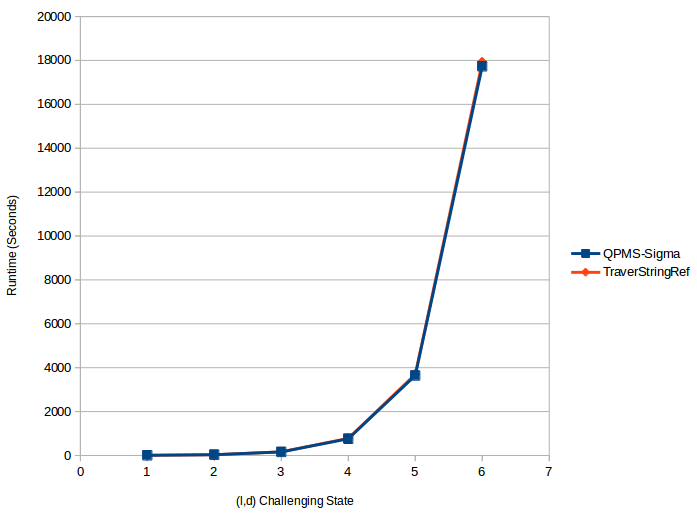}
	\caption{Run Time Comparison between qPMS-Sigma vs TraverStringRef in the challenging states for q=10}
	\label{fig:chart2}
\end{figure}

% Citation examples
%\input{citeexamples.tex}

% Another chapter
%\input{anotherchapter.tex}

% Chapter showing example of index creation
%\input{indexcreation.tex}

% conclusion
\chapter{Conclusion}\label{conclusion}

\section{Summary of Our Work}
The main task of our thesis is to develop an effective algorithm for the Planted Motif Search problem. Our proposed algorithm qPMS-Sigma tries to find the $ (l, d) $-motifs for $ n $ given sequences. It is an exact version of the PMS algorithm. qPMS-Sigma is based on the previous PMS algorithms qPMSPrune, qPMS7, TraverStringRef and PMS8. In our proposed algorithm we introduce clever techniques to compress the input sequences and thus space complexity is improved. We also include a faster comparison technique of the $ l $-mers by adding bitwise comparison techniques. As we know the PMS is a NP-Hard problem, it is exponential in terms of time. So, parallel implementation techniques are proposed at the end of our work.

\section{Future Prospects of Our Work}
Though in our thesis we have given some ideas about parallel implementation of our algorithm, it is not implemented and tested on multiprocessor system. All comparison
of our algorithm with the stated algorithms are done in single processor system. To understand the real speedup and slackness we need to experiment in a system involving a network of nodes having multiple processors. Our work is the extension of PMS8 codes which involves openMPI libraries. In future we want to continue our work using OpenMPI project which is a open source Message Passing Interface. Apart from the parallelism, more advanced pruning conditions can be used to reduce the search space. Different randomized approaches can also be included for improving the runtime, though it will make the exact version of our algorithm approximate.

% Bibliographies and appendices
% You do not need to change anything in this file. If you want to
% change the reference style, comment/uncomment the \bibliographystyle
% lines

\clearpage
\renewcommand\bibname{References}
\addcontentsline{toc}{chapter}{References}

% Comment/uncomment as suits you
 \bibliographystyle{ieeetr} %% IEEE transaction style
\bibliography{buetcseugthesis}

% Index, comment this out if you do not want to create an index
\printindex

\appendix
% Algorithms
%\input{algorithms.tex}

% Codes
% Code settings
\lstset{
  language=C++, % C, C++, Java, SQL are from the around hundred available
  basicstyle=\ttfamily,
  numbers=left,
  numberstyle=\footnotesize,
  stepnumber=1, 
  numbersep=2.0mm,
  breaklines=true
}

\chapter{Codes}\label{ch:codes}

\section{Compressed String}\label{string_compress}

Here is a sample C++ structure for compressing a 1000 character long DNA string

\begin{lstlisting}
//letterToInt[i] means a unique id for each character of the alphabet;
//Here, 
//letterToInt['A']=0;
//letterToInt['B']=1;
//letterToInt['C']=2;
//letterToInt['D']=3;

struct CompressedlMer
{
    unsigned int **cs;//[2][32];
    int l;
    CompressedlMer(char *string, int start, int l)    {//constructor
        int group = 0;    
        int intPerGroup = ceil(l/sizeof(unsigned int));
        int temp = l;
        while(temp)    {
            group++;
            temp = temp/2;
        }
        cs = new unsigned int*[group];
        for(int i = 0; i < group; i++)
            cs[i] = new unsigned int[intPerGroup];
            
        for(int i = 0; i < l; i++)    {
            int j = letterToInt[string[start+i]];
            int positionMask = 1<<(i%sizeof(unsigned int));
            int intNo = i/sizeof(unsigned int);
            int grp = 0;
            while(j>0)    {
                if(j&1) cs[grp][intNo] = cs[grp][intNo] | positionMask;
                j = j/2;
                grp++;
            }
        }
    }
    
    ~CompressedlMer()    {//destructor
        for(int i = 0,j=1; j < l; i++,j*=2)    {
            delete [] cs[i];
        }
        delete [] cs;
    }
};

int hammingDist(const CompressedlMer &sa, const CompressedlMer &sb)    {
    int intPerGroup = ceil(sa.l/sizeof(unsigned int));
    int hamDist = 0;
    for(int i = 0; i < intPerGroup; i++)    {
        unsigned int flag = ~(0);    //all bits are set to 1
        for(int group = 0, j = 1; j < sa.l; group++, j*=2)    {
            flag = flag | (sa.cs[group][i]^sb.cs[group][i]);
        }
        hamDist = hamDist + __builtin_popcount(flag);
    }
    return hamDist;
}
\end{lstlisting}

\end{document}